\documentclass{pasj00}

\newcommand{\nh}{\ensuremath{N_{\rm{H}}}}
\newcommand{\nodata}{~$\cdots$~}
\newcommand{\lndata}{\leavevmode \leaders \hrule depth \dimexpr0.4pt-0.7ex height 0.7ex \hfill \kern0pt}

\newcommand{\lbb}{\ensuremath{L_{\rm{BB}}}}
\newcommand{\ktbb}{\ensuremath{k_{\rm{B}}T}_{BB}}
\newcommand{\kttp}{\ensuremath{k_{\rm{B}}T}_{TH}}
\newcommand{\ktnl}{\ensuremath{k_{\rm{B}}T}_{NL}}
\newcommand{\vem}{\ensuremath{EM}}
\newcommand{\ars}{\ensuremath{A_{\rm{RS}}}}
\newcommand{\nnl}{\ensuremath{N_{\rm{NL}}}}
\newcommand{\lxh}{\ensuremath{L_{\rm{HX}}}}
\newcommand{\fxh}{\ensuremath{F_{\rm{HX}}}}
\newcommand{\lxs}{\ensuremath{L_{\rm{SX}}}}
\newcommand{\fxs}{\ensuremath{F_{\rm{SX}}}}

\SetRunningHead{D. Takei et al.}{X-ray Development of V2672 Ophiuchi}

\Received{2013/07/09}
\Accepted{2013/11/21}
\Published{}

\begin{document}
\title{X-ray Development of the Classical Nova V2672 Ophiuchi with Suzaku}
\author{
Dai \textsc{Takei},\altaffilmark{1,2}
Masahiro \textsc{Tsujimoto},\altaffilmark{3}
Jeremy J. \textsc{Drake},\altaffilmark{2}
and Shunji \textsc{Kitamoto}\altaffilmark{4}
}
\altaffiltext{1}{Institute of Physical and Chemical Research (RIKEN),
                 RIKEN SPring-8 Center, \\
		 1-1-1 Kouto, Sayo, Hyogo 679-5148, Japan}
		 \email{takei@spring8.or.jp}
\altaffiltext{2}{Smithsonian Astrophysical Observatory (SAO),
		 60 Garden Street, Cambridge, MA 02138, USA}
\altaffiltext{3}{Japan Aerospace Exploration Agency (JAXA),
		 Institute of Space and Astronautical Science (ISAS), \\
		 3-1-1 Yoshino-dai, Chuo-ku, Sagamihara, Kanagawa 252-5210, Japan}
\altaffiltext{4}{Department of Physics, Rikkyo University,
                 3-34-1 Nishi-Ikebukuro, Toshima, Tokyo 171-8501, Japan}

\maketitle
\begin{abstract}
 We report the Suzaku detection of a rapid flare-like X-ray flux amplification early in
 the development of the classical nova V2672 Ophiuchi. Two target-of-opportunity
 $\sim$25~ks X-ray observations were made 12 and 22 days after the outburst. The flux
 amplification was found in the latter half of day 12. Time-sliced spectra are
 characterized by a growing supersoft excess with edge-like structures and a relatively
 stable optically-thin thermal component with K$\alpha$ emission lines from highly
 ionized Si. The observed spectral evolution is consistent with a model that has a
 time development of circumstellar absorption, for which we obtain the decline rate
 of $\sim$10--40\%\ in a time scale of 0.2~d on day 12. Such a rapid drop of absorption
 and short-term flux variability on day 12 suggest inhomogeneous ejecta with dense
 blobs/holes in the line of sight. Then on day 22 the fluxes of both supersoft and
 thin-thermal plasma components become significantly fainter. Based on the serendipitous
 results we discuss the nature of this source in the context of both short- and
 long-term X-ray behavior.
\end{abstract}

\section{Introduction}\label{introduction}
Classical novae are violent stellar explosions in close binaries that contain a white
dwarf primary accreting material from a red dwarf or giant secondary. Sudden hydrogen
fusion is triggered by a thermonuclear runaway in the accreted gas when it exceeds a
critical limit \citep{starrfield2008a}. The general picture of nova evolution is
characterized by the development of photospheric emission that dominates first in the
optical, and then in supersoft X-rays ($\lesssim$1~keV) as the surrounding ejecta become
less opaque. For detailed reviews of novae see e.g., \citet{warner2003a,bode2008a}.

The spectral transition from optical to supersoft X-rays is not always smooth as the
simple generalized picture predicts. X-ray snapshots with Swift (e.g., \cite{schwarz2011s})
have discovered large and rapid amplitude variability, particularly early in the
development of the transition: e.g., in RS~Oph~2006 \citep{osborne2011t}, V458~Vul
\citep{ness2009s}, Nova~LMC~2009a \citep{bode2009d}, and KT~Eri \citep{beardmore2010l}.
Short-term changes in supersoft X-ray emission can potentially provide valuable insights
into the structure and nature of the emitting region, and the spectral development of
the source must also play a key part in the physical interpretation.

The mechanism underlying short-term changes in nova outburst X-ray fluxes is still not
understood, and neither is it known whether similar physical processes are at work in
different systems. \citet{drake2003}
observed a short 15~min X-ray burst from V1494~Aql during the supersoft source phase
that remains unexplained. \citet{ness2003} suggested white dwarf spin is related to the
22~min oscillations seen in V4743~Sgr. \citet{osborne2011t} suggested absorption due to
clumpy ejecta for the large flickering of RS~Oph 2006. \citet{beardmore2010s} found a
binary period with a 1.77~d modulation in CSS\,081007:030559$+$054715. \citet{ness2012f}
argued that reforming accretion blobs caused a flux instability in U~Sco~2010. The
present set of observations are still sparse, and further progress in this direction
demands additional X-ray observations. Since they are currently difficult to predict,
progress is likely to come from capturing short-term transient phenomena serendipitously.

\medskip
The purpose of this paper is to report the Suzaku detection of a rapid flare-like X-ray
flux amplification early in the development of the classical nova V2672 Ophiuchi (V2672
Oph). This represents a rare addition to the currently very small sample of rapid X-ray
flux change on short timescales for classical novae, and the advent of time-sliced
X-ray spectroscopy provides a unique opportunity to investigate its behavior in detail.

The plan of this paper is as follows: In section~\ref{target} we summarize previous
studies of this target. In section~\ref{observe} we introduce our target-of-opportunity
observations obtained with the Suzaku satellite. In section~\ref{analysis} we describe
the analysis and results. In section~\ref{discussion} we discuss the results in the
context of the nature of the source, and a summary is presented in section~\ref{summary}.

\section{Target (V2672 Ophiuchi)}\label{target}
\subsection{Ground-Based Observations}

\begin{figure}[tb]
 \begin{center}
  \FigureFile(85mm,85mm){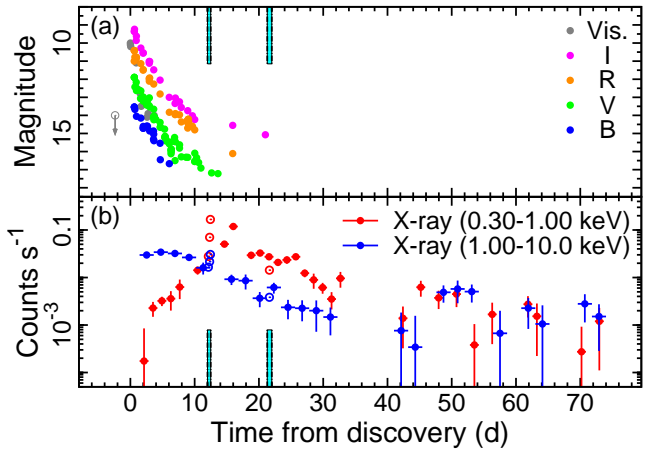}
 \end{center}
 \vspace{-3.000mm}
 \caption{Development of (a) optical and (b) X-ray brightness in the nova outburst of
 V2672 Oph. The origin of the abscissa in modified Julian date is 55059.52~d, when
 the nova event was first discovered by \citet{nakano2009a}. The times of the Suzaku
 observations are indicated by the cyan regions. (a) \textit{B}-, \textit{V}-, \textit{R}-,
 \textit{I}-band, and visual (i.e., non-filtered) magnitudes are shown color-coded. The
 upper-limit of the visual magnitude before the discovery of the nova is indicated by the
 open circle with the downward arrow. Optical data are from the American Association of
 Variable Star Observers (AAVSO), the Variable Star Observers League in Japan (VSOLJ),
 and telegrams by \citet{nakano2009a,nakano2009b,nissinen2009,ayani2009b,munari2009}.
 No ground-based photometry was taken after about day 20 as the optical brightness
 faded quickly. (b) Background-subtracted Swift/XRT count rates in the 0.3--1.0~keV and
 1.0--10~keV energy bands are shown color-coded. Open circles indicate our time-averaged
 Suzaku results in each phase that were converted into Swift/XRT count rates using the
 \textit{pimms} software. The Swift observations were conducted by G.\,Schwarz on behalf
 of the Swift Nova-CV group \citep{schwarz2009,schwarz2011s}.
 }\label{fg:lcurve_optical}
\end{figure}

An optical nova was discovered on 2009 August 16.515~UT (modified Julian date
55059.52~d) in the constellation Ophiuchus at (RA, Dec) $=$ ($\timeform{17h38m19.68s}$,
$\timeform{-26D44'14.0''}$) in the equinox J2000.0 \citep{nakano2009a}. Throughout this
paper we define the epoch of the discovery as the origin of time. The nova event showed
a visual magnitude of 10.0~mag at discovery in unfiltered CCD camera photometry
\citep{nakano2009a}. Nothing brighter than an \textit{R}-band magnitude of 20.8~mag was
found at the nova position in the Digitized Sky Survey images taken from 1991 to 1996
\citep{nakano2009a,nakano2009b}, suggesting that the source suddenly brightened by more
than $\sim$10~mag.  The event was identified as a classical nova explosion, and was
named Nova Ophiuchi 2009 and V2672 Ophiuchi \citep{nakano2009b}.

Ground-based photometric and spectroscopic observations were subsequently conducted
\citep{nakano2009a,nakano2009b,nissinen2009,ayani2009a,ayani2009b,munari2009,munari2011a}.
Figure~\ref{fg:lcurve_optical}a shows the development of the nova brightness in optical and
infrared wavelengths.  Assuming the maximum \textit{V}-band magnitude of 11.9~mag on 2009
August 17.134~UT \citep{nakano2009a}, decline rates $t_{2}$ and $t_{3}$ are $\sim$3 and
4~d, where $t_{2}$ and $t_{3}$ are the times to fade by 2 and 3~mag from its optical
maximum, respectively; based on this the nova type is classified as extremely fast. The
\textit{B}--\textit{V} color was estimated to be 1.81~mag on 2009 August 17.47~UT
\citep{munari2009}. An optical spectrum exhibited H$\alpha$, H$\beta$,
O\emissiontype{I}, and possibly He\emissiontype{I} emission lines in the early phase
\citep{munari2009}.  The velocity width of the H$\alpha$ line was $\sim$8000~km~s$^{-1}$
on day 1 \citep{ayani2009a}. The rapid decline of its brightness and the extremely high
velocity of ejecta suggest that the binary system comprises a massive white dwarf
\citep{munari2009}.  The distance to V2672 Oph was estimated to be 19$\pm$2~kpc based on
the empirical relations of \citet{downes2000,cohen1988n,schmidt1957d} using the optical
data of \citet{munari2011a}. In the radio regime, synchrotron emission with a flat
spectral index ($\alpha$ $\lesssim$ 1.2) was detected on day 15.61 with the C
configuration of the Very Large Array, suggesting that high energy electrons were
produced in shocked ejecta \citep{krauss2009}.

\subsection{Space-Based Observations}
In the pre-outburst phase, no significant emission was found either in X-ray archival
images taken by the ROSAT and ASCA satellites, or in the XMM-Newton slew survey
\citep{schwarz2009}. After the discovery, intensive monitoring observations were
conducted by Swift and INTEGRAL \citep{schwarz2009,schwarz2011s}.  X-rays were first
detected 1.43 days after the nova discovery using Swift, with a count rate of
0.017~s$^{-1}$.  Ultraviolet emission was also detected with Swift, while no
$\gamma$-ray emission was found with INTEGRAL 4 and 7 days after the discovery.
Figure~\ref{fg:lcurve_optical}b shows the development of Swift X-ray count rates. The
turn-on and turn-off timescales of the supersoft phase were estimated to be 22$\pm$2
and 28$\pm$2~d, respectively, based on the Swift data \citep{schwarz2011s}. We
requested target-of-opportunity observations with the Suzaku satellite to perform deep
X-ray spectroscopy early in the evolution of V2672 Oph.

\section{Observation and Reduction}\label{observe}

\begin{table}[tb]
 \begin{center}
  \caption{Suzaku observations of V2672 Oph.}\label{tb:observation}
  \vspace{-1.00mm}
  \begin{tabular}{ll|ll}
   \hline\hline
   \multicolumn{2}{l|}{Observation ID}                      & 904002010        & 904002020        \\
   $\textit{t}$\footnotemark[$*$]                    & (d)  & 12.2             & 21.6             \\
   $\textit{t}\rm{_{start}}$\footnotemark[$\dagger$] & (UT) & 2009-08-28 12:20 & 2009-09-06 19:38 \\
   $\textit{t}\rm{_{end}}$\footnotemark[$\dagger$]   & (UT) & 2009-08-29 00:00 & 2009-09-07 11:30 \\
   $\textit{t}\rm{_{exp}}$\footnotemark[$\ddagger$]  & (ks) & 23.1             & 25.1             \\
   \hline
   \multicolumn{3}{@{}l@{}}{\hbox to 0pt{\parbox{85mm}{\footnotesize
   \par\noindent
   \footnotemark[$*$] Elapsed days in the middle of each observation from the discovery
   of V2672 Oph (55059.52~d in the modified Julian date).
   \par\noindent
   \footnotemark[$\dagger$] Start and end dates of the Suzaku observations.
   \par\noindent
   \footnotemark[$\ddagger$] Net exposure times averaged over the three operating CCDs.
   }\hss}}
  \end{tabular}
 \end{center}
\end{table}

Target-of-opportunity observations of V2672 Oph were performed using the Suzaku X-ray
observatory twice, on 2009 August 28 and September 6 (12 and 22 days after the
discovery, respectively; see table~\ref{tb:observation}).  Suzaku has two X-ray
instruments in operation \citep{mitsuda2007}: the X-ray Imaging Spectrometer (XIS;
\cite{koyama2007x}) and the Hard X-ray Detector (HXD; \cite{takahashi2007,kokubun2007}).
In this paper, we concentrate on the XIS data in which significant X-ray emission from
V2672 Oph was detected.  The HXD signal was seriously contaminated by neighboring hard
X-ray sources.

The XIS is equipped with four X-ray CCDs at the foci of four co-aligned X-ray telescope
modules \citep{serlemitsos2007t}. Three of them (XIS0, 2, and 3) are front-illuminated
(FI) CCDs sensitive in the 0.4--12~keV energy band, while the remaining one (XIS1) is a
back-illuminated (BI) CCD sensitive in the 0.2--12~keV range. Each chip has a format of
1024$\times$1024 pixels and covers a $\sim$18$\arcmin\times$18$\arcmin$ field of view.
Two radioactive $\atom{Fe}{}{55}$ sources illuminate two corners of each CCD. Prior to
the observation dates, XIS2 and a part of XIS0 became dysfunctional and the data
were excluded. The absolute energy scale is accurate to $\lesssim$10~eV, and the full
width at half maximum energy resolutions are $\sim$180 and 230~eV at 5.9~keV for the FI
and BI CCDs, respectively. The XIS was operated in the normal clocking mode with an 8~s
frame time. The observations were aimed to put V2672 Oph at the center of the XIS field
with almost the same roll angle in each.

Data were processed with the standard pipeline version 2.4\footnote{See
http://www.astro.isas.jaxa.jp/suzaku/process/ for details.}, in which events were
removed during South Atlantic anomaly passages, when night-earth elevation angles were
below 5$^\circ$, and day-earth elevation angles were below 20$^\circ$. The net exposure
times are 23 and 25~ks on days 12 and 22, respectively.  For data reduction and
analysis, we used the High Energy Astrophysics software package version 6.7, the
calibration database version xis20090925 and xrt20080709, and the Sherpa fitting program
\citep{freeman2001s} in the Chandra Interactive Analysis of Observations package
\citep{fruscione2006c}.

\section{Analysis}\label{analysis}
\subsection{Image Analysis}\label{analysis_image}

\begin{figure}[tb]
 \begin{center}
  \FigureFile(85mm,85mm){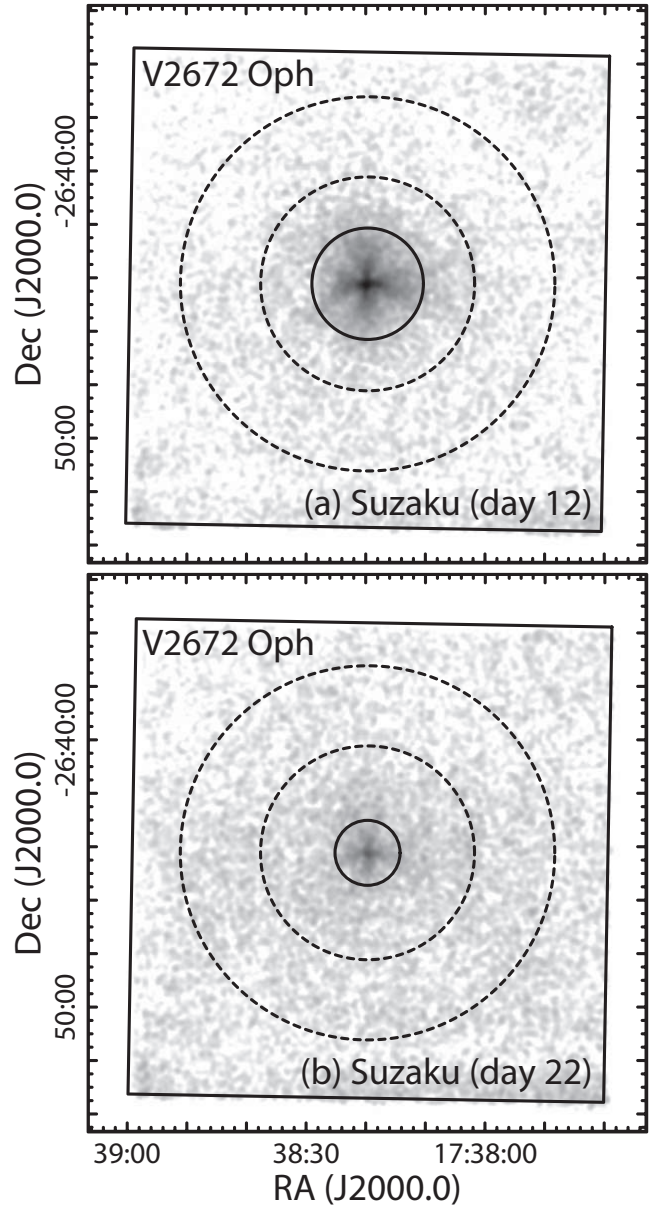}
 \end{center}
 \vspace{-3.000mm}
 \caption{Smoothed XIS images (a) 12 and (b) 22 days after the discovery. Events taken
 with the three CCDs in the 0.2--12.0~keV energy band were merged in each observation.
 The source and background extraction regions are indicated with the solid circles and
 the dashed annuli, respectively.
 }\label{fg:image_xis}
\end{figure}

Figure~\ref{fg:image_xis} shows smoothed XIS images in the 0.2--12.0~keV energy band on
days 12 and 22. Events taken with each XIS were merged.  Astrometry of the XIS images was
registered by matching the position of V2672~Oph with that by an optical observation of
\citet{nakano2009a}.  For temporal and spectral analysis, source events were accumulated
from a circular region with radii of 120 and 70 pixels (2\farcm1 and 1\farcm2) that were
adaptively chosen to maximize the signal-to-noise ratio on days 12 and 22, respectively.
Background events were accumulated from annular regions with inner and outer radii of
4$\arcmin$ and 7$\arcmin$ in each observation.

\subsection{Temporal Analysis}\label{analysis_timing}

\begin{figure*}[tb]
 \begin{center}
  \FigureFile(85mm,85mm){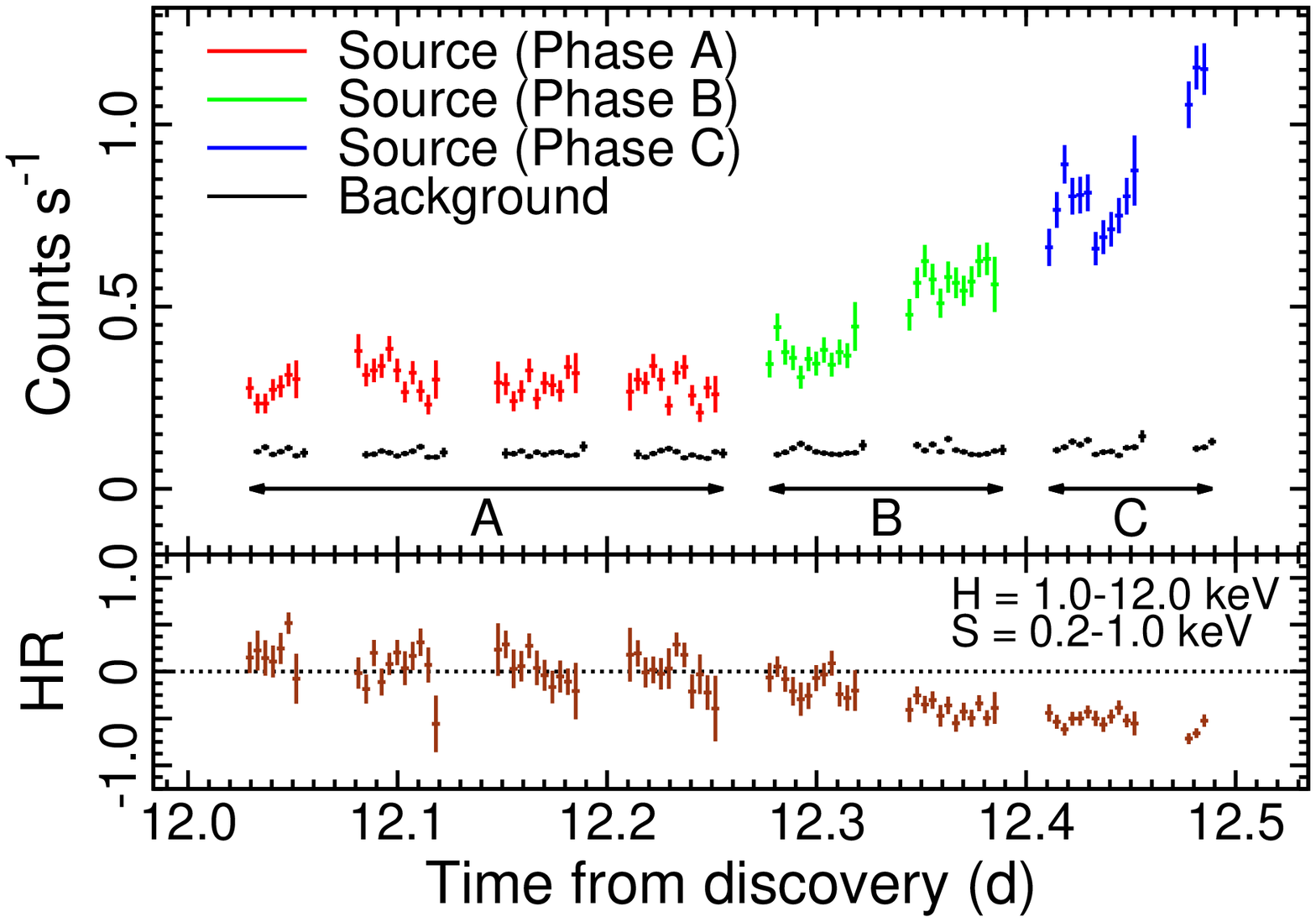}
  \FigureFile(85mm,85mm){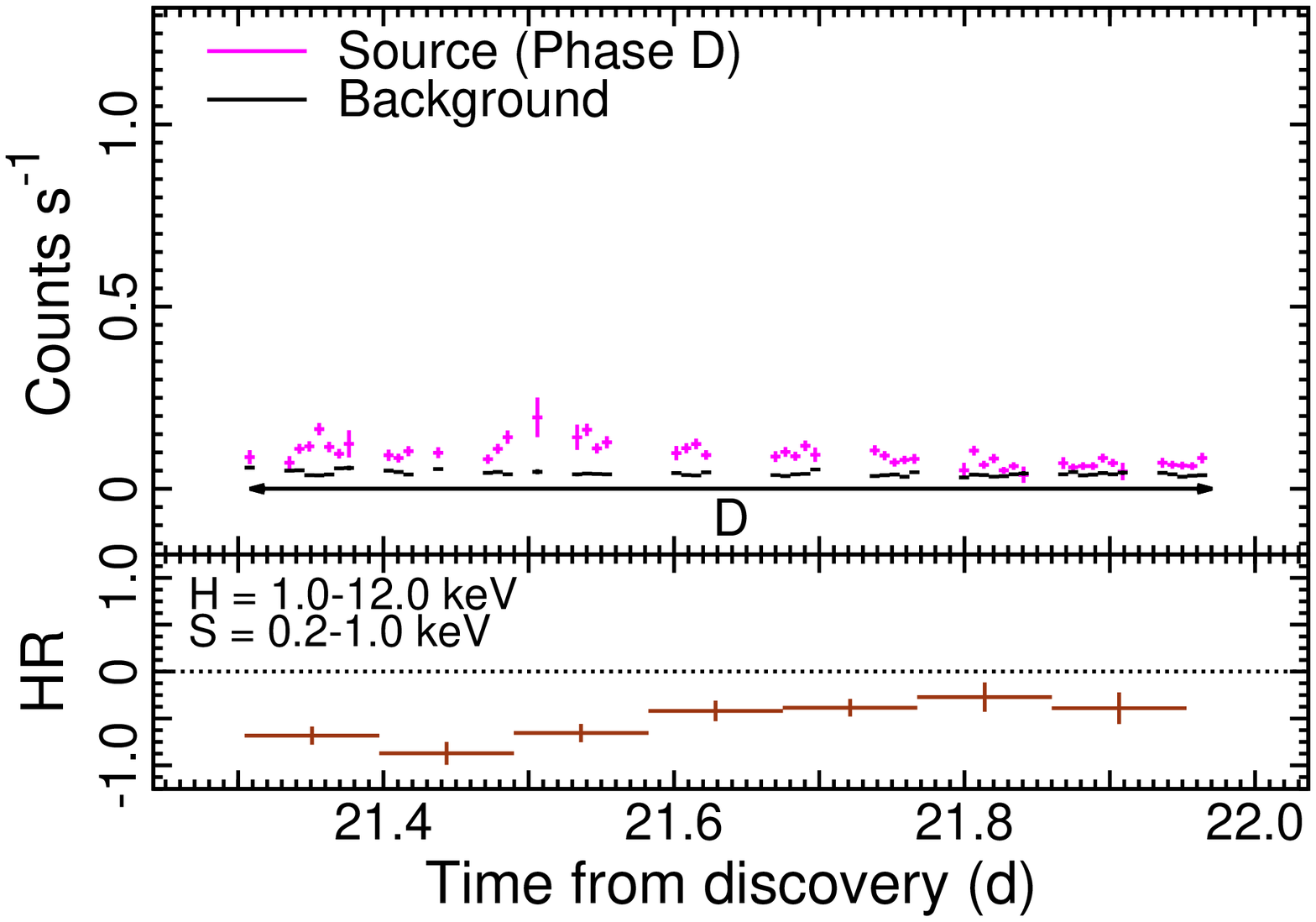}
 \end{center}
 \vspace{-3.000mm}
 \caption{Top: Source and background XIS light curves on days 12 and 22. Events recorded
 by the three CCDs in the 0.2--12.0~keV energy band were merged. Source count rates are
 shown color-coded in each time-slice and are labelled in terms of behavioral phase (see
 text). Background count rates are normalized by the ratio of extraction areas of source
 and background events. The normalized background count rates in the two observations
 are quite similar and their difference is almost accounted for by the respective sizes
 of the source extraction regions. Bottom: The hardness ratios (HRs) defined by
 (H$-$S)/(H$+$S), where H and S are rates in the 0.2--1.0~keV and 1.0--12.0~keV energy
 bands, respectively. Data on day 22 were re-binned because of the poor photon statistics.
 }\label{fg:lcurve_xis}
\end{figure*}

Figure~\ref{fg:lcurve_xis} shows source and background light curves in the 0.2--12.0~keV
energy band. On day 12, while the X-ray flux level had been stable in the first half of
the observation, a prominent flux increase by a factor of nearly five was seen in the
latter half. Significant correlation was found between the flux increase and the
spectral softening on day 12 (Figure~\ref{fg:lcurve_xis}). On day 22 the flux was
relatively stable, showing variations,
though at a much lower flux level. According to the flux development, we divided data on
day 12 into three time slices (phases A--C), while data on day 22 are treated separately
as phase D. The times defining each phase are as follows.

\begin{description}
 \item[Phase A :] day 12.0--12.3 (before the rise)
 \item[Phase B :] day 12.3--12.4 (early in the rise)
 \item[Phase C :] day 12.4--12.5 (later in the rise)
 \item[Phase D :] day 21.3--22.0 (full data set on day 22)
\end{description}

For the sake of comparison, we further converted time-averaged fluxes in each phase into
Swift/XRT count rates by the \textit{pimms} software \citep{mukai1993p}. The long-term
light curve including the Suzaku data (figure~\ref{fg:lcurve_optical}) shows a signature
of large fluctuations below 1.0~keV around day 12--16.

\subsection{Spectral Analysis}\label{analysis_spectra}

\begin{figure}[tb]
 \begin{center}
  \FigureFile(85mm,85mm){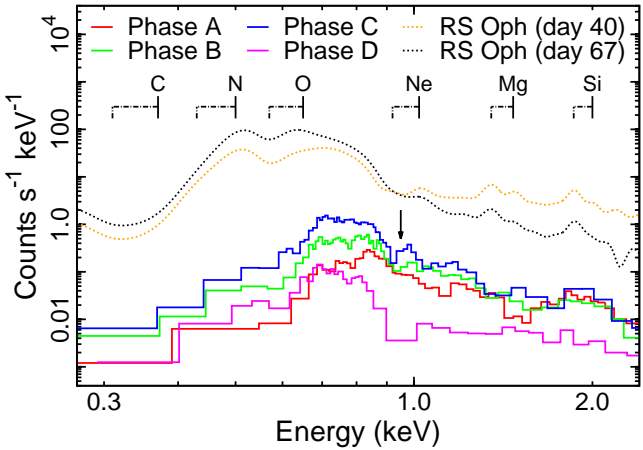}
 \end{center}
 \vspace{-3.000mm}
 \caption{Development of the X-ray spectra of V2672 Oph. The background-subtracted BI
 spectra are shown color-coded.  The energies of plausible K$\alpha$ emission lines are
 labeled with solid-and-dashed lines for H- and He-like ions, respectively. The downward
 arrow indicates the unidentified structure described in section \ref{analysis_spectra}.
 For comparison, the dashed curves are X-ray spectra of RS~Oph 2006 taken by the Chandra
 grating instruments 40 and 67 days after the outburst, convolved with the Suzaku XIS-BI
 response. 
 }\label{fg:spxis}
\end{figure}

Figures~\ref{fg:spxis} and \ref{fg:spectrum_mod} show background-subtracted XIS spectra
of each time-slice. The two FI spectra with nearly identical responses were merged,
while the BI spectra were treated separately for spectral analysis. The detector and
mirror responses were generated by the $\texttt{xisrmfgen}$ and $\texttt{xissimarfgen}$
tools \citep{ishisaki2007}, respectively.

The X-ray spectra are characterized by a soft excess below $\sim$1~keV and a hard tail up
to $\sim$10~keV. On day 12 the soft component was highly variable, which corresponds to
the flare-like brightening, while the hard component was relatively stable. K$\alpha$
emission from Si\emissiontype{XIII} was clearly present and stronger than
that from Si\emissiontype{XIV}. Prominent structures were found at both 0.6 and 0.9~keV,
possibly by ionized absorption similar to those found in other novae (e.g.,
\cite{ness2009h}). In addition an unidentified structure was seen at 0.9--1.0~keV in
phase~C (the downward arrow in figure~\ref{fg:spxis}); this cannot be explained by
K$\alpha$ emission lines from highly ionized Ne assuming their rest energies. It also
seems unlikely that either the Ne\emissiontype{IX} or Ne\emissiontype{X} emission lines
are shifted by the Doppler effect with a velocity of $\gtrsim$10000~km~s$^{-1}$. One
possible explanation might be a pseudo flux peak between prominent Ne\emissiontype{IX}
and Ne\emissiontype{X} absorption lines, though no such structure has been reported in
previous studies of novae. Then on day 22, both soft and hard components had faded
significantly. Over time during this period dominant emission shifted toward lower
energies.

\subsubsection{Basic Modeling}\label{analysis_spec_basic}
We first applied a combination of blackbody and optically-thin collisionally-ionized
thermal plasma components to model softer and harder parts, respectively, of the
V2672~Oph spectra. This approach has been used before for many other novae (e.g.,
\cite{hernanz2002c,ness2007x,takei2008d}), and is driven by expectations
of optically-thin emission from expanding ejecta and optically-thick photospheric
emission from the supersoft source. The MEKAL code \citep{mewe1985c} was adopted to
represent the optically-thin plasma emission, and elemental abundances for this were
fixed to be solar. Both emission components were multiplied with an interstellar
absorption model (TBabs; \cite{wilms2000}). Free parameters in the combined model
were a hydrogen-equivalent column density ($\nh$), blackbody temperature ($\ktbb$)
and bolometric luminosity ($\lbb$), and MEKAL plasma temperature ($\kttp$) and X-ray
volume emission measure ($\vem$). We also calculated X-ray fluxes ($\fxs$ and $\fxh$)
and absorption-corrected luminosities ($\lxs$ and $\lxh$) in softer and harder
(0.2--1.0 and 1.0--10~keV) energy bands, respectively. For the luminosity estimates
the distance to V2672~Oph was assumed to be 19~kpc \citep{munari2011a}. The best-fit
parameters producing minima in the Xspec version of the $\chi^{2}$ statistic with
data variance are summarized in table~\ref{tb:par}.
This basic model provides a good representation of the data above $\sim$1~keV, while
significant residuals were found in the softer part of the spectrum, particularly
near 0.6, 0.9, and 0.9--1.0~keV, where we infer the presence of significant additional
local structures.

Based on the basic modeling, we then measured $\chi^{2}$ increases with an additional
null hypothesis that either $\nh$, $\ktbb$, or $\lbb$ was constant during the rapid
flux rise in phases A--C. The best-fit values for comparison are also summarized in
table~\ref{tb:par}. The resulting $\chi^{2}$ values and the degrees of freedom indicate
that F-test statistics for changing $\nh$, $\ktbb$, or $\lbb$ are 2.48, 0.63, and 0.60,
which can be converted to F-test null hypothesis probabilities of 0.09, 0.53, and 0.55,
respectively. Even though these statistical tests are not strictly accurate in our case
because of non-normality (see e.g., \cite{orlandini2012b}), it seems likely that the
time development in phases A--C would be subject to a change in absorption.

We emphasize here that the blackbody approximation is only suitable as a first order
description of the underlying spectrum, and the best-fit effective temperature and
luminosity can differ greatly from actual values, whether the spectral shape is
well-fitted or not (e.g., \cite{krautter1996,takei2010}).  This is because nova
supersoft X-ray spectra result from radiative transfer within an expanding medium
and exhibit numerous complex absorption features, often combined with emission lines
(e.g., \cite{ness2007s,ness2009h,ness2012f}). While it might be possible to obtain a
statistically more acceptable fit from adding further local components to the model,
the exercise would be arbitrary in the context of attempting to learn more about the
emitting source. In the following subsections we focus on the softer part and adopt
more sophisticated spectral models in the place of the blackbody component.

\subsubsection{Atmosphere Modeling}\label{analysis_spec_sssrauch}
As the second fitting approach we applied a Non-Local Thermodynamic Equilibrium (NLTE)
atmospheric model (T\"{u}bingen NLTE Model-Atmosphere Fluxes; e.g., \cite{rauch1997i}),
instead of a blackbody model for the soft component. The theoretical spectral energy
distributions are calculated for elements H--Ca with solar abundances, and approximated
formulae are employed to treat Stark line broadening \citep{rauch1993n}. Assuming a
massive white dwarf close to the Chandrasekhar limit (e.g., \cite{munari2009}), we adopt
a logarithmic surface gravity $\log{g}$~$=$~9. The effective temperature ($\ktnl$) and a
normalization parameter of the model flux ($\nnl$) are treated as free parameters. The
best-fit results are also summarized in table~\ref{tb:par}. The use of these atmospheric
models slightly improves the goodness of fit, in comparison to the basic blackbody model.

\subsubsection{Comparison with RS Ophiuchi 2006}\label{analysis_spec_comp}

\begin{figure}[tb]
 \begin{center}
  \FigureFile(85mm,85mm){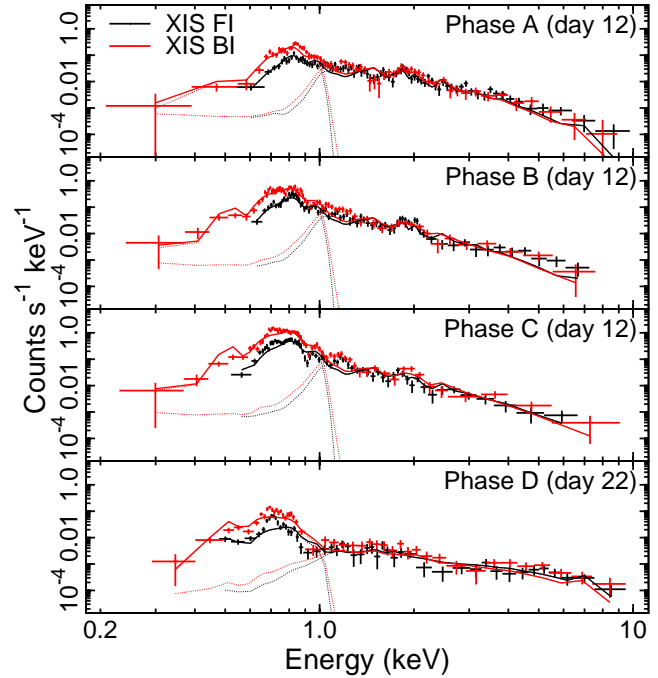}
 \end{center}
 \vspace{-3mm}
 \caption{Background-subtracted XIS spectra and the best-fit models using the normalized
 supersoft component of RS Oph 2006 taken on day 67 and the MEKAL plasma code convolved
 with photoelectric absorption.  The best-fit models are shown by the solid lines, while
 each component is illustrated by dashed lines.
 }\label{fg:spectrum_mod}
\end{figure}

To go one step further with spectroscopy, we compared the X-ray spectra of V2672~Oph
with those of RS Oph 2006, which is one of only a handful of novae for which detailed
multi-wavelength data sets including X-rays have been obtained to date. X-ray grating
spectra of the supersoft emission in RS Oph 2006 were obtained at three different
epochs (days 54 [XMM-Newton], 40 and 67 [Chandra]; e.g., \cite{ness2009h}). In order to
compare the medium-resolution V2672~Oph CCD spectra with high-resolution grating data
for RS~Oph 2006, we first calculated photon flux spectra using instrumental responses
of each grating observation. The resulting spectra were then convolved with the XIS
responses, assuming that instrumental line broadening of the grating spectra is
negligible; the resolving power is more than 10 times higher than that of the CCD
spectrometers. Figure~\ref{fg:spxis} shows these resultant degraded grating spectra
of RS~Oph 2006, which were obtained in the supersoft X-ray phase, 40 and 67 days after
the outburst using Chandra. We here focused on the Chandra data set because the X-ray
spectra at the two different epochs were obtained by the same instrument. The spectral
development of V2672 Oph is basically analogous to that of RS Oph 2006, in which the
peak energy of the supersoft component shifts toward lower energies as time progresses
(see figure~\ref{fg:spxis}).  The heavy attenuation at 0.9~keV, which we attributed to
atmospheric absorption, is also similar to a pattern seen in RS Oph 2006 spectra (see
also figure~\ref{fg:spxis}).

We fitted the V2672 Oph spectra by a combination of the MEKAL plasma component and the
unfolded RS~Oph 2006 spectrum, convolved with a photoelectric absorption (phabs in
XSPEC). Grating data in the 0.2--1.0~keV energy band that were dominated by supersoft
emission with negligible contribution ($\lesssim$~1\%) of the plasma component of RS~Oph
2006 were used for fitting. The normalization of the RS~Oph 2006 component ($\ars$)
was treated as a free parameter. Elemental abundances for the photoelectric absorption
model were fixed to be solar. We compared the goodness of fit in each phase using
Chandra data at the two different epochs. Empirically, the day 67 model provides the
best fits in all phases. The best-fit results are listed in table~\ref{tb:par} and are 
shown in figure~\ref{fg:spectrum_mod}.

\begin{table*}[htbp]
 \begin{center}
  \setlength{\tabcolsep}{3.00mm}
  \caption{Best-fit parameters of the X-ray spectra of V2672 Oph.}\label{tb:par}
  \vspace{-1.00mm}
  \begin{small}
  \begin{tabular}{lllcccc}
   \hline\hline
   Comp.        & Par.                               & Unit                                & Phase A\footnotemark[$*$] & Phase B\footnotemark[$*$] & Phase C\footnotemark[$*$] & Phase D\footnotemark[$*$] \\
   \hline
   \multicolumn{7}{c}{TBabs $\times$ [Blackbody $+$ MEKAL] (section~\ref{analysis_spec_basic})} \\
   \hline
   Absorption   & $\nh$                              & (10$^{22}$ cm$^{-2}$)               & 2.08$_{-0.15}^{+0.16}$    & 1.61$_{-0.27}^{+0.14}$    & 1.59$_{-0.15}^{+0.15}$    & 0.69$_{-0.15}^{+0.17}$    \\
   Blackbody    & $\ktbb$                            & (eV)                                & 44.3$_{-2.24}^{+2.48}$    & 48.8$_{-2.57}^{+5.75}$    & 47.7$_{-2.64}^{+3.97}$    & 60.8$_{-5.38}^{+5.99}$    \\
                & $\lbb$\footnotemark[$\S$]          & (10$^{40}$ erg~s$^{-1}$)            & 104.8$_{-69.8}^{+276.0}$  & 10.0$_{-6.95}^{+21.8}$    & 24.1$_{-17.8}^{+71.2}$    & 0.003$_{-0.002}^{+0.013}$ \\
   Plasma       & $\kttp$                            & (keV)                               & 0.97$_{-0.07}^{+0.08}$    & 1.02$_{-0.11}^{+0.36}$    & 1.00$_{-0.16}^{+0.33}$    & 5.04$_{-1.64}^{+6.28}$    \\
                & $\vem$\footnotemark[$\S$]          & (10$^{57}$ cm$^{-3}$)               & 12.7$_{-1.82}^{+1.00}$    & 8.97$_{-2.78}^{+1.75}$    & 10.5$_{-2.87}^{+2.92}$    & 0.79$_{-0.13}^{+0.16}$    \\
   \hline
   Flux         & $\fxs$\footnotemark[$\dagger$]     & (10$^{-12}$ erg s$^{-1}$ cm$^{-2}$) & 0.55$_{-0.55}^{+0.03}$    & 1.45$_{-1.45}^{+0.08}$    & 2.87$_{-2.85}^{+0.04}$    & 0.38$_{-0.38}^{+0.07}$    \\
                & $\fxh$\footnotemark[$\ddagger$]    & (10$^{-12}$ erg s$^{-1}$ cm$^{-2}$) & 0.62$_{-0.11}^{+0.03}$    & 0.67$_{-0.18}^{+0.03}$    & 0.83$_{-0.26}^{+0.04}$    & 0.24$_{-0.06}^{+0.03}$    \\
   Luminosity   & $\lxs$\footnotemark[$\dagger\S$]   & (10$^{40}$ erg s$^{-1}$)            & $\sim$30.2                & $\sim$3.39                & $\sim$8.86                & $\sim$0.002               \\
                & $\lxh$\footnotemark[$\ddagger\S$]  & (10$^{35}$ erg s$^{-1}$)            & 4.65$_{-0.83}^{+0.21}$    & 2.87$_{-0.77}^{+0.14}$    & 4.22$_{-1.30}^{+0.19}$    & 0.14$_{-0.03}^{+0.02}$    \\
   \hline
   \multicolumn{3}{l}{$\chi^{2}/\rm{d.o.f}$}                                               & 239/144                   & 429/138                   & 893/129                   & 522/92                    \\
   \hline\hline
   \multicolumn{7}{c}{TBabs $\times$ [Blackbody $+$ MEKAL] (section~\ref{analysis_spec_basic}, for comparison)} \\
   \hline
   Absorption   & $\nh$                              & (10$^{22}$ cm$^{-2}$)               & \multicolumn{3}{c}{\lndata\, 1.71$_{-0.10}^{+0.08}$ \lndata}                      & \nodata                   \\
   Blackbody    & $\ktbb$                            & (eV)                                & 50.9$_{-1.73}^{+2.24}$    & 47.0$_{-1.44}^{+1.86}$    & 45.6$_{-1.35}^{+1.77}$    & \nodata                   \\
                & $\lbb$\footnotemark[$\S$]          & (10$^{40}$ erg~s$^{-1}$)            & 3.81$_{-2.24}^{+4.14}$    & 25.0$_{-14.7}^{+27.1}$    & 71.1$_{-42.2}^{+78.4}$    & \nodata                   \\
   Plasma       & $\kttp$                            & (keV)                               & 1.25$_{-0.09}^{+0.11}$    & 0.99$_{-0.11}^{+0.12}$    & 0.96$_{-0.18}^{+0.13}$    & \nodata                   \\
                & $\vem$\footnotemark[$\S$]          & (10$^{57}$ cm$^{-3}$)               & 8.16$_{-0.89}^{+0.79}$    & 9.89$_{-1.52}^{+1.43}$    & 11.9$_{-1.99}^{+3.73}$    & \nodata                   \\
   \hline
   \multicolumn{3}{l}{$\chi^{2}/\rm{d.o.f}$}                                               & \multicolumn{3}{c}{\lndata\, 1579/413 \lndata}                                    & \nodata                   \\
   \hline\hline
   Absorption   & $\nh$                              & (10$^{22}$ cm$^{-2}$)               & 1.92$_{-0.10}^{+0.09}$    & 1.70$_{-0.09}^{+0.08}$    & 1.63$_{-0.09}^{+0.08}$    & \nodata                   \\
   Blackbody    & $\ktbb$                            & (eV)                                & \multicolumn{3}{c}{\lndata\, 46.9$_{-1.37}^{+1.65}$ \lndata}                      & \nodata                   \\
                & $\lbb$\footnotemark[$\S$]          & (10$^{40}$ erg~s$^{-1}$)            & 25.8$_{-14.6}^{+28.4}$    & 25.5$_{-14.3}^{+27.5}$    & 35.1$_{-19.5}^{+37.2}$    & \nodata                   \\
   Plasma       & $\kttp$                            & (keV)                               & 1.00$_{-0.07}^{+0.09}$    & 0.99$_{-0.12}^{+0.11}$    & 0.99$_{-0.15}^{+0.13}$    & \nodata                   \\
                & $\vem$\footnotemark[$\S$]          & (10$^{57}$ cm$^{-3}$)               & 11.2$_{-1.41}^{+1.38}$    & 9.92$_{-1.46}^{+1.43}$    & 11.0$_{-1.83}^{+1.79}$    & \nodata                   \\
   \hline
   \multicolumn{3}{l}{$\chi^{2}/\rm{d.o.f}$}                                               & \multicolumn{3}{c}{\lndata\, 1565/413 \lndata}                                    & \nodata                   \\
   \hline\hline
   Absorption   & $\nh$                              & (10$^{22}$ cm$^{-2}$)               & 1.93$_{-0.09}^{+0.09}$    & 1.72$_{-0.09}^{+0.08}$    & 1.60$_{-0.09}^{+0.08}$    & \nodata                   \\
   Blackbody    & $\ktbb$                            & (eV)                                & 46.7$_{-1.39}^{+1.62}$    & 46.8$_{-1.40}^{+1.63}$    & 47.4$_{-1.44}^{+1.69}$    & \nodata                   \\
                & $\lbb$\footnotemark[$\S$]          & (10$^{40}$ erg~s$^{-1}$)            & \multicolumn{3}{c}{\lndata\, 27.7$_{-15.3}^{+30.2}$ \lndata}                      & \nodata                   \\
   Plasma       & $\kttp$                            & (keV)                               & 1.00$_{-0.07}^{+0.09}$    & 0.98$_{-0.13}^{+0.11}$    & 1.00$_{-0.14}^{+0.14}$    & \nodata                   \\
                & $\vem$\footnotemark[$\S$]          & (10$^{57}$ cm$^{-3}$)               & 11.3$_{-1.39}^{+1.38}$    & 10.0$_{-1.46}^{+1.46}$    & 10.7$_{-1.81}^{+1.79}$    & \nodata                   \\
   \hline
   \multicolumn{3}{l}{$\chi^{2}/\rm{d.o.f}$}                                               & \multicolumn{3}{c}{\lndata\, 1565/413 \lndata}                                    & \nodata                   \\
   \hline\hline
   \multicolumn{7}{c}{TBabs $\times$ [NLTE Atmosphere $+$ MEKAL] (section~\ref{analysis_spec_sssrauch})} \\
   \hline
   Absorption   & $\nh$                              & (10$^{22}$ cm$^{-2}$)               & 2.17$_{-0.05}^{+0.05}$    & 1.90$_{-0.05}^{+0.04}$    & 1.75$_{-0.04}^{+0.04}$    & 1.46$_{-0.07}^{+0.04}$    \\
   Atmosphere   & $\ktnl$                            & (eV)                                & 69.1$_{-0.03}^{+0.04}$    & 69.1$_{-0.03}^{+0.04}$    & 69.1$_{-0.03}^{+0.04}$    & 68.9$_{-1.61}^{+0.04}$    \\
                & $\nnl$\footnotemark[$\|$]          & (10$^{-3}$)                         & 1.99$_{-0.48}^{+0.55}$    & 1.71$_{-0.41}^{+0.47}$    & 1.91$_{-0.43}^{+0.49}$    & 0.08$_{-0.03}^{+0.02}$    \\
   Plasma       & $\kttp$                            & (keV)                               & 1.03$_{-0.08}^{+0.17}$    & 1.01$_{-0.10}^{+0.13}$    & 1.02$_{-0.11}^{+0.22}$    & 3.33$_{-1.58}^{+4.63}$    \\
                & $\vem$\footnotemark[$\S$]          & (10$^{57}$ cm$^{-3}$)               & 11.6$_{-1.35}^{+1.36}$    & 9.82$_{-1.44}^{+1.44}$    & 10.4$_{-1.94}^{+1.75}$    & 1.11$_{-0.21}^{+0.22}$    \\
   \hline
   Flux         & $\fxs$\footnotemark[$\dagger$]     & (10$^{-12}$ erg s$^{-1}$ cm$^{-2}$) & 0.56$_{-0.06}^{+0.04}$    & 1.68$_{-0.15}^{+0.09}$    & 3.91$_{-0.30}^{+0.16}$    & 0.52$_{-0.06}^{+0.03}$    \\
                & $\fxh$\footnotemark[$\ddagger$]    & (10$^{-12}$ erg s$^{-1}$ cm$^{-2}$) & 0.65$_{-0.05}^{+0.04}$    & 0.68$_{-0.05}^{+0.04}$    & 0.85$_{-0.08}^{+0.06}$    & 0.22$_{-0.04}^{+0.03}$    \\
   Luminosity   & $\lxs$\footnotemark[$\dagger\S$]   & (10$^{40}$ erg s$^{-1}$)            & $\sim$4.68                & $\sim$4.01                & $\sim$4.51                & $\sim$0.18                \\
                & $\lxh$\footnotemark[$\ddagger\S$]  & (10$^{35}$ erg s$^{-1}$)            & 5.68$_{-0.40}^{+0.32}$    & 4.91$_{-0.40}^{+0.32}$    & 5.60$_{-0.51}^{+0.39}$    & 0.26$_{-0.05}^{+0.03}$    \\

   \hline
   \multicolumn{3}{l}{$\chi^{2}/\rm{d.o.f}$}                                               & 194/144                   & 197/138                   & 346/129                   & 282/92                    \\
   \hline\hline
   \multicolumn{7}{c}{phabs $\times$ [Unfolded RS Oph 2006 $+$ MEKAL] (section~\ref{analysis_spec_comp})} \\
   \hline
   Absorption   & $\nh$                              & (10$^{22}$ cm$^{-2}$)               & 1.22$_{-0.06}^{+0.06}$    & 0.92$_{-0.04}^{+0.04}$    & 0.82$_{-0.03}^{+0.03}$    & 0.38$_{-0.04}^{+0.04}$    \\
   RS Oph 2006  & $\ars$                             & (RS Oph 2006 on day 67)             & 0.87$_{-0.20}^{+0.26}$    & 0.63$_{-0.10}^{+0.12}$    & 0.90$_{-0.11}^{+0.13}$    & 0.010$_{-0.002}^{+0.003}$ \\
   Plasma       & $\kttp$                            & (keV)                               & 1.24$_{-0.21}^{+0.08}$    & 1.22$_{-0.11}^{+0.09}$    & 1.30$_{-0.12}^{+0.12}$    & 4.91$_{-1.37}^{+2.67}$    \\
                & $\vem$\footnotemark[$\S$]          & (10$^{57}$ cm$^{-3}$)               & 6.82$_{-0.49}^{+0.90}$    & 6.02$_{-0.43}^{+0.44}$    & 6.70$_{-0.52}^{+0.53}$    & 0.74$_{-0.09}^{+0.10}$    \\
   \hline
   Flux         & $\fxs$\footnotemark[$\dagger$]     & (10$^{-12}$ erg s$^{-1}$ cm$^{-2}$) & 0.59$_{-0.14}^{+0.18}$    & 1.74$_{-0.29}^{+0.34}$    & 4.03$_{-0.51}^{+0.57}$    & 0.52$_{-0.11}^{+0.13}$    \\
                & $\fxh$\footnotemark[$\ddagger$]    & (10$^{-12}$ erg s$^{-1}$ cm$^{-2}$) & 0.60$_{-0.04}^{+0.08}$    & 0.64$_{-0.05}^{+0.05}$    & 0.78$_{-0.06}^{+0.06}$    & 0.23$_{-0.03}^{+0.03}$    \\
   Luminosity   & $\lxs$\footnotemark[$\dagger\S\#$] & (10$^{37}$ erg s$^{-1}$)            & $>$1.16                   & $>$0.90                   & $>$1.36                   & $>$0.01                   \\
                & $\lxh$\footnotemark[$\ddagger\S$]  & (10$^{35}$ erg s$^{-1}$)            & 0.49$_{-0.04}^{+0.07}$    & 0.43$_{-0.03}^{+0.03}$    & 0.48$_{-0.04}^{+0.04}$    & 0.070$_{-0.008}^{+0.009}$ \\

   \hline
   \multicolumn{3}{l}{$\chi^{2}/\rm{d.o.f}$}                                               & 265/145                   & 252/139                   & 417/130                   & 289/93                    \\
   \hline\hline
   \multicolumn{7}{@{}l@{}}{\hbox to 0pt{\parbox{180mm}{\footnotesize
   \par\noindent
   \footnotemark[$*$] Statistical uncertainties indicate the 90\% confidence ranges in
   the Xspec version of the $\chi^{2}$ statistic.
   \par\noindent
   \footnotemark[$\dagger$] Values are in the 0.2--1.0~keV energy band.
   \par\noindent
   \footnotemark[$\ddagger$] Values are in the 1.0--10~keV energy band.
   \par\noindent
   \footnotemark[$\S$] Values are for a distance of 19~kpc \citep{munari2011a}.
   \par\noindent
   \footnotemark[$\#$] Values are the normalization of astrophysical fluxes at a
   distance of 1~pc.
   \par\noindent
   \footnotemark[$\|$] Values are lower limits because of the additional unresolved
   absorption in the RS Oph 2006 component.
   }\hss}}
  \end{tabular}
  \end{small}
 \end{center}
\end{table*}

\section{Discussion}\label{discussion}
\subsection{Supersoft Emission}\label{dis_soft}
The supersoft component is similar in character to that of other novae, and this
can be considered photospheric emission powered by post-outburst residual nuclear
burning on the white dwarf. The long-term supersoft X-ray rise clearly coincides
with the optical decline until about day 15 (figure~\ref{fg:lcurve_optical}),
suggesting that dominant emission shifted toward higher energies as the mass outflow
diminished and consequently the photon escape region was formed deeper in the ejected
material. The temperatures of 40--70~eV are also typical of those found in other
supersoft sources and novae (e.g., \cite{greiner2000c}). Then the supersoft emission
faded out on day 22, suggesting that nuclear burning rate was reduced on the white
dwarf surface.

It has to be noted that an actual X-ray luminosity is difficult to estimate because
of large spectral model-dependent systematic uncertainties in the supersoft emission.
Also the distance estimate of 19~kpc is subject to uncertainties \citep{munari2011a}.
The blackbody and NLTE atmosphere luminosity estimates are almost certainly too high
because they exceed the Eddington limit of a Chandrasekhar white dwarf, while the
RS~Oph 2006 values are underestimated as the model itself contains intrinsic
absorption. Within the large uncertainty from the three model approaches we chose as
a working hypothesis values of luminosity of the NLTE atmosphere model. The values
at 19~kpc are 10$^{39}$--10$^{40}$~~erg~s$^{-1}$ in the 0.2--1.0~keV energy band. The
softer band luminosity estimates clearly exceed the Eddington values.

The rapid flux rise on day 12 is consistent with a model in which the intrinsic
absorption of the nova outburst declined with time. This is because the absorption
dropped with increasing supersoft X-ray intensity (table~\ref{tb:par}). In addition,
the hydrogen-equivalent absorption column densities on day 12 significantly exceeded
the interstellar Galactic value of 3.8$\times$10$^{21}$~cm$^{-2}$, which is from
H\emissiontype{I} intensity maps \citep{kalberla2005t} for a 1$\arcdeg$ cone radius
at the V2672~Oph position and can be considered an upper limit. The F-test results
might further support this scenario. This suggests that the supersoft emission would
be affected by the time development of surrounding ejecta.

In contrast to the absorber, no significant growth can be otherwise confirmed in the
emission components during the rapid flux rise on day 12. This indicates that the
broad-band fitting of the limited CCD data does not allow us to examine whether the
supersoft component contributes to the spectral changes simultaneously. We here
cannot exclude a scenario that the time development of emission lines like those of
RS~Oph~2006 (e.g., \cite{nelson2008,ness2009h}) contributes to the spectral changes.
Further, the nature of unidentified structure at 0.9--1.0~keV in phase~C is still an
open question. Higher resolution spectroscopy would have been required to provide a
reliable diagnostic of the time development of the emission components.

Assuming that the intrinsic absorption contributed to the rapid flux rise on day 12,
the decline rate of the column densities poses a challenge to understanding the time
development of the mass outflow. For example, assuming spherically symmetric ejecta
with a gradient density distribution, the gas density profile is
\begin{equation}
 \rho \sim \frac{1}{4\pi{}r^{2}}\frac{\dot{M}}{v}, \label{eq:rho}
\end{equation}
where $r$ is a radial distance, $v$ is an ejecta velocity, and $\dot{M}$ is a
time-dependent mass-loss rate. The intrinsic absorption can be approximated by
\begin{eqnarray}
 \nh \sim \int_{r_{wd}}^{r_{out}}{\rho{dr}} \sim \frac{\dot{M}}{4\pi{}vr_{wd}}\left(1 - \frac{r_{wd}}{r_{out}}\right), \label{eq:nhe}
\end{eqnarray}
where $r_{wd}$ is a white dwarf radius and $r_{out}$ is the outermost radius of
ejecta. Assuming that the time scale of the terminal velocity decline is much
larger than the time for the ejecta to reach the outermost radius, $r_{out}$
$\approx$ $vt$ at time $t$. By adopting a constant velocity assuming an initial
free expansion stage, $r_{out}$ $\gg$ $r_{wd}$ and the intrinsic absorption
decreases only in proportion to the decrease in $\dot{M}$. The time-dependent
mass-loss rate can be further assumed as (e.g., \cite{bode2008a,takei2013x})
\begin{equation}
 \dot{M} = \dot{M}_{0}(t_{0}/t)^{p}, \label{eq:mdot}
\end{equation}
where $\dot{M}_{0}$ is an initial value normalized at time $t$ $=$ $t_{0}$ $=$
1~s and $p$ is a measure of the speed of its decline. This indicates that we
can make a crude estimate of $p$ from the decline rate of the observed column
densities.

The difference in the absorption column densities between days 12 and 22 is
0.6--1.7$\times$10$^{22}$~cm$^{-2}$, which corresponds to a decline rate of
about 30--80\%\ in a time scale of 10~d. This does not depend on the fitting
models and is much too fast to be explained by thinning of uniform ejecta,
even assuming a non-decelerated expansion velocity. Based on the result,
a measure of a decline rate given by equation \ref{eq:mdot} is $p$ $\sim$
0.6--2.7, suggesting that the long-term $\dot{M}$ decline time scale is
similar to that derived from U Sco 2010 \citep{takei2013x}. The value is also
consistent with a prediction that $p$ is of order unity (e.g., \cite{bode2008a}).
In contrast, the difference in the short-term column densities between phases
A and C on day 12 is 2--8$\times$10$^{21}$~cm$^{-2}$, and its decline rate is
about 10--40\%\ in a time scale of 0.2~d relative to the phase A values. This
can be converted to $p$ $\sim$ 6.3--31, according to equation \ref{eq:mdot}.
Though we cannot exclude a scenario that the reality has more complex trends
of $\dot{M}$, the values indicate that the cause of the rapid rise on day 12
cannot reasonably be explained by the same framework of the long-term development.

We finally expect that the rapid absorption drop on day 12 and the subsequent
large fluctuations (figure~\ref{fg:lcurve_optical}) are likely to betray a
non-uniform density distribution of the ejected material (e.g., see figure 2
in \cite{shaviv2010} and figure 5 in \cite{williams2013n}). Less dense regions
of the ejecta would be preferentially broken by the strong radiation pressure,
in particular since the photospheric emission is close to the Eddington
luminosity (e.g., \cite{owocki1988,shaviv2010}). The surrounding material would
be porous and clumpy, and full of holes like a Swiss cheese. The relatively
stable light curve in phase A can be interpreted as the presence of a dense blob
or thin hole in the line of sight, while the flux rise in phases B--C results
from the hole expanding, or the blob gradually moving out of the line of sight.

\subsection{Optically-Thin Thermal Plasma Emission}\label{dis_hard}
The Swift light curve in the harder energy band exhibits a continuous flux decline,
except for a brightening around day 50, since its eruption (figure~\ref{fg:lcurve_optical}),
indicating that the origin of the harder component is different from that of the
supersoft component. This is similar to the findings from previous nova explosions
(e.g., \cite{tsujimoto2009}). The temperature of this component is several keV,
indicating an origin in shock-excited plasma resulting from kinematic interactions
within the outflow (e.g., \cite{mukai2001}). The harder band estimates are typical of
those derived for other novae (e.g., \cite{tsujimoto2009}). The difference of plasma
temperatures in phases A--C and D might, in principle, be of interest for constraining a
heating mechanism, though the values have large uncertainties because of the poor photon
statistics and the high background levels in the higher energy band.

\section{Summary}\label{summary}
We conducted target-of-opportunity observations of the classical nova V2672 Ophiuchi, 12
and 22 days after the discovery, using the Suzaku X-ray satellite. The X-ray light curve
exhibited a clear flux amplification on day 12, then fluxes became significantly smaller
on day 22. X-ray spectra have been modeled by a supersoft source and an optically-thin
thermal plasma component convolved with a photoelectric absorption model. The spectral
evolution of the X-ray rise can be accounted for a reduction in absorption over time,
with a decline of 10--40\%\ on a time scale of 0.2~d. This rapid drop of absorption and
the associated short-term variability on day 12 indicate inhomogeneous and porous ejecta,
with dense blobs or holes in the line of sight.

\bigskip

The authors appreciate the reviewer for useful suggestions. We thank the Suzaku telescope
managers for allocating a part of the director's discretionary time, and also thank Greg
Schwarz, the principal investigator of the Swift observations, and the Nova-CV group for
providing the Swift data. Optical data are from AAVSO and VSOLJ databases. This research
has made use of data obtained from: (1) DARTS by C-SODA in JAXA/ISAS, (2) HEASARC by
NASA/GSFC, and (3) SIMBAD by CDS, France. We acknowledge Jan-Uwe Ness for his comments
on this paper. D.\,T. acknowledges financial support from JSPS and by the Special
Postdoctoral Researchers Program in RIKEN. J.\,J.\,D. was supported by the NASA contract
NAS8-03060 to the CXC and thanks the Director, H.~Tananbaum, for continuing advice and
support.

\bibliographystyle{ms}
\bibliography{ms}

\end{document}